\documentclass[twocolumn]{aastex63}
\graphicspath{{./}{figures/}}%https://www.overleaf.com/project/5ff275f63fb6925a33c80984
\usepackage{natbib}
\usepackage{xcolor}
\usepackage{amsmath}

\begin{document}

\title{A Search for the 3.5 keV Line from the Milky Way's Dark Matter Halo with HaloSat}

\author{E.M. Silich}
\affil{Department of Physics and Astronomy, University of Iowa, Iowa City, IA 52242, USA}
\author{K. Jahoda}
\affil{NASA Goddard Space Flight Center, Greenbelt, MD 20771, USA}
\author{L. Angelini}
\affil{NASA Goddard Space Flight Center, Greenbelt, MD 20771, USA}
\author{P. Kaaret}
\affil{Department of Physics and Astronomy, University of Iowa, Iowa City, IA 52242, USA}
\author{A. Zajczyk}
\affil{Department of Physics and Astronomy, University of Iowa, Iowa City, IA 52242, USA}
\affil{NASA Goddard Space Flight Center, Greenbelt, MD 20771, USA}
\affil{Center for Space Sciences and Technology, University of Maryland, Baltimore County, 1000 Hilltop Circle, Baltimore, MD 21250, USA}
\author{D.M. LaRocca}
\affil{Department of Physics and Astronomy, University of Iowa, Iowa City, IA 52242, USA}
\affil{Department of Astronomy and Astrophysics, 525 Davey Labs, Pennsylvania State University, University Park, PA 16802, USA}
\author{R. Ringuette}
\affil{Department of Physics and Astronomy, University of Iowa, Iowa City, IA 52242, USA}
\affil{NASA Goddard Space Flight Center, Greenbelt, MD 20771, USA}
\affil{Adnet Systems, Bethesda, MD 20817, USA}
\author{J. Richardson}
\affil{Department of Physics and Astronomy, University of Iowa, Iowa City, IA 52242, USA} 

\begin{abstract}
Previous detections of an X-ray emission line near 3.5 keV in galaxy clusters and other dark matter-dominated objects have been interpreted as observational evidence for the decay of sterile neutrino dark matter. Motivated by this, we report on a search for a 3.5 keV emission line from the Milky Way's galactic dark matter halo with HaloSat. As a single pixel, collimated instrument, HaloSat observations are impervious to potential systematic effects due to grazing incidence reflection and CCD pixelization, and thus may offer a check on possible instrumental systematic errors in previous analyses. We report non-detections of a $\sim$3.5 keV emission line in four HaloSat observations near the Galactic Center. In the context of the sterile neutrino decay interpretation of the putative line feature, we provide \textcolor{black}{90\% confidence level upper limits on the 3.5 keV line flux and 7.1 keV sterile neutrino mixing angle: $F \leq 0.077$ ph cm$^{-2}$ s$^{-1}$ sr$^{-1}$ and $\sin^2(2\theta) \leq 4.25 \times 10^{-11}$. The HaloSat mixing angle upper limit was calculated using a modern parameterization of the Milky Way's dark matter distribution, and in order to compare with previous limits, we also report the limit calculated using a common historical model.} The HaloSat mixing angle upper limit places constraints on a number of previous mixing angle estimates derived from observations of the Milky Way's dark matter halo and galaxy clusters, and excludes several previous detections of the line. The upper limits cannot, however, entirely rule out the sterile neutrino decay interpretation of the 3.5 keV line feature.
\end{abstract}


\section{Introduction}
While there exists strong observational evidence for the existence of dark matter based on studies of galaxies and galaxy clusters (e.g. \citet{1933AcHPh...6..110Z, 1937ApJ....86..217Z, 1980ApJ...238..471R, 2006ApJ...648L.109C}), the composition of dark matter is unknown. The Standard Model does not contain a viable dark matter particle candidate, which has motivated extensions of the Standard Model such as the $\nu$ Minimal Standard Model ($\nu$MSM; \citet{2005PhLB..620...17A,2005PhLB..631..151A}) that propose hypothetical particle candidates. The dark matter particle candidate of interest to this work is the sterile neutrino, which is a right-handed counterpart to the left-handed active neutrino \citep{2005PhLB..631..151A,2005PhLB..620...17A,2012PDU.....1..136B,2019PrPNP.104....1B}. The sterile neutrino is so named since it is a $SU(2)$-singlet particle that does not experience weak interactions; it's interactions are limited to mixing with the active ($SU(2)$-doublet) neutrinos \citep{2001PhRvD..64b3501A}. 

The keV-scale sterile neutrino should spontaneously decay at a rate of

\begin{equation}
    \! \Gamma_{\gamma} (m_s, \theta) = 1.38 \times 10^{-29} \,\rm{s}^{-1} \left( \frac{\sin^2(2\theta)}{10^{-7}}\right) \left( \frac{m_s}{1 \; \rm{keV}}\right)^5
\end{equation}
for $m_s$ the mass of the sterile neutrino and $\theta$ the mixing angle between the active and sterile states \citep{1982PhRvD..25..766P}. This radiative decay process should produce an active neutrino and a photon with energy $E = m_s/2$. For a keV-scale sterile neutrino, the decay should produce an emission line that is observable by modern X-ray observatories in dark-matter dominated objects like galaxies and galaxy clusters \citep{2001PhRvD..64b3501A,2001ApJ...562..593A}, with the strength of the line determined by the sterile-active neutrino mixing angle. 

The first potential observational evidence for sterile neutrino dark matter decay was presented in \citet{2014ApJ...789...13B} as a detection of an unidentified X-ray emission line near 3.5 keV in \textit{XMM-Newton} observations of 73 galaxy clusters. After determining no obvious astrophysical origin, the 3.5 keV line was interpreted as a decay signature of a $\sim$7 keV sterile neutrino, since the decay of a $\sim$7 keV sterile neutrino should produce an X-ray photon at $E \sim 3.5$ keV. \citet{2014ApJ...789...13B} reported a corresponding 7.1 keV sterile neutrino mixing angle of $\sin^2(2\theta) \sim 7 \times 10^{-11}$. 

This detection was followed up by a series of observations of dark-matter dominated objects with CCD instruments. \citet{2014PhRvL.113y1301B} detected a line near 3.5 keV with $\sim$4.4$\sigma$ significance and a mixing angle of $\sin^2(2\theta) = (2-20) \times 10^{-11}$ using \textit{XMM-Newton} observations of the M31 galaxy and the Perseus galaxy cluster. \citet{2016ApJ...831...55B} used observations of 47 galaxy clusters from \textit{Suzaku} that resulted in a non-detection of a $\sim$3.5 keV line with a mixing angle upper limit of $\sin^2(2\theta) \leq 6.1 \times 10^{-11}$. Since then, detections and non-detections of the line that are inconsistent with one another have introduced controversy as to the reality of the 3.5 keV line feature.

\citet{2018ApJ...854..179C} searched the spectrum of the Cosmic X-ray Background with \textit{Chandra} and reported a line detection at $\sim$2.5$-3 \sigma$ significance with a mixing angle of $\sin^2(2\theta) = (0.83-2.75) \times 10^{-10}$. However, \citet{2020ApJ...905..146S} recently observed the Milky Way's dark matter halo with $\sim$51 Ms of \textit{Chandra} observations resulting in a non-detection and a mixing angle upper limit of $\sin^2(2\theta) \leq  2.58 \times 10^{-11}$, in disagreement with \citet{2018ApJ...854..179C}.

\citet{2018arXiv181210488B} observed the Milky Way Galactic Center with \textit{XMM-Newton} and detected the 3.5 keV line at local significances of $\sim$2.1$-5 \sigma$ and mixing angles of $\sin^2(2\theta) = (1.6-2.1) \times 10^{-11}$. Also using \textit{XMM-Newton}, \citet{2020MNRAS.497..656B} searched for the line in the spectra of 117 galaxy clusters and reported a non-detection and mixing angle upper limit of $\sin^2(2\theta) \leq 4.4 \times 10^{-11}$.

\citet{2020Sci...367.1465D} studied \textit{XMM-Newton} blank-sky observations looking through the Milky Way's dark matter halo, and reported a non-detection of a 3.5 keV line corresponding to a mixing angle upper limit \textcolor{black}{of $\sin^2(2\theta) \sim 10^{-12}$. This limit is} inconsistent with the decaying dark matter particle interpretation of the line. However, the \citet{2020Sci...367.1465D} results have been recently criticized for several aspects of their analysis, which placed overestimated constraints on the dark matter decay rate (see \citet{2020arXiv200406601B} and \citet{2020arXiv200406170A}).

Only a few non-CCD instruments have performed systematic searches for the line. \citet{2016PhRvD..94l3504N} reported a line near 3.5 keV at $\sim$11$\sigma$ significance using deep sky observations of the Milky Way's dark matter halo with \textit{NuSTAR}, but the line is near the lower bound of \textit{NuSTAR}’s sensitivity, where large uncertainty in the response may be present. \textit{Hitomi} observations of the Perseus galaxy cluster also resulted in a non-detection and corresponding flux upper limit \citep{2017ApJ...837L..15A} that was inconsistent with the presence of a 3.5 keV line at the strength reported by \citet{2014ApJ...789...13B}. This inconsistency was attributed to a systematic error in the \textit{XMM-Newton} result. In particular, \citet{2014ApJ...789...13B} and \citet{2017ApJ...837L..15A} note that at \textit{XMM-Newton}'s CCD spectral resolution, even a 1\% variation in the effective area curve could introduce an artifact that might be interpreted as a faint line-like feature. It is therefore important to investigate the presence of the 3.5 keV line using experiments with systematics that are different from one another in order to ensure the faint emission line is not associated with an effect inherent to one particular type of instrument. 

The HaloSat CubeSat is an all-sky survey that observed in the $0.4 - 7$ keV energy range from 2018 October to 2021 January. HaloSat was designed primarily to study diffuse $\sim$0.6 keV emission expected in the diffuse halo of the Milky Way galaxy (\citet{2019ApJ...884..162K}; \citet{2020NatAs...4.1072K}). Due to the large field of view (FoV; $\sim$100 deg$^{2}$), HaloSat has a large grasp \footnote{The grasp is defined in \citet{2019ApJ...884..162K} as the product of the HaloSat effective area with the FoV.} and good sensitivity to diffuse emission. The energy resolution is comparable to the CCD experiments on \textit{Chandra} and \textit{XMM-Newton}, and the response is simple and well understood \citep{2020JATIS...6d4005Z} so that HaloSat observations of diffuse emission from the Milky Way, which fills the FoV, have comparable sensitivity to the large observatories. Any systematics in the response are different than the systematics associated with grazing incidence reflection and the complex charge transfer and multi-pixel effects inherent in CCD observations.

\begin{figure*}[t]
    \epsscale{1.0}
    \plotone{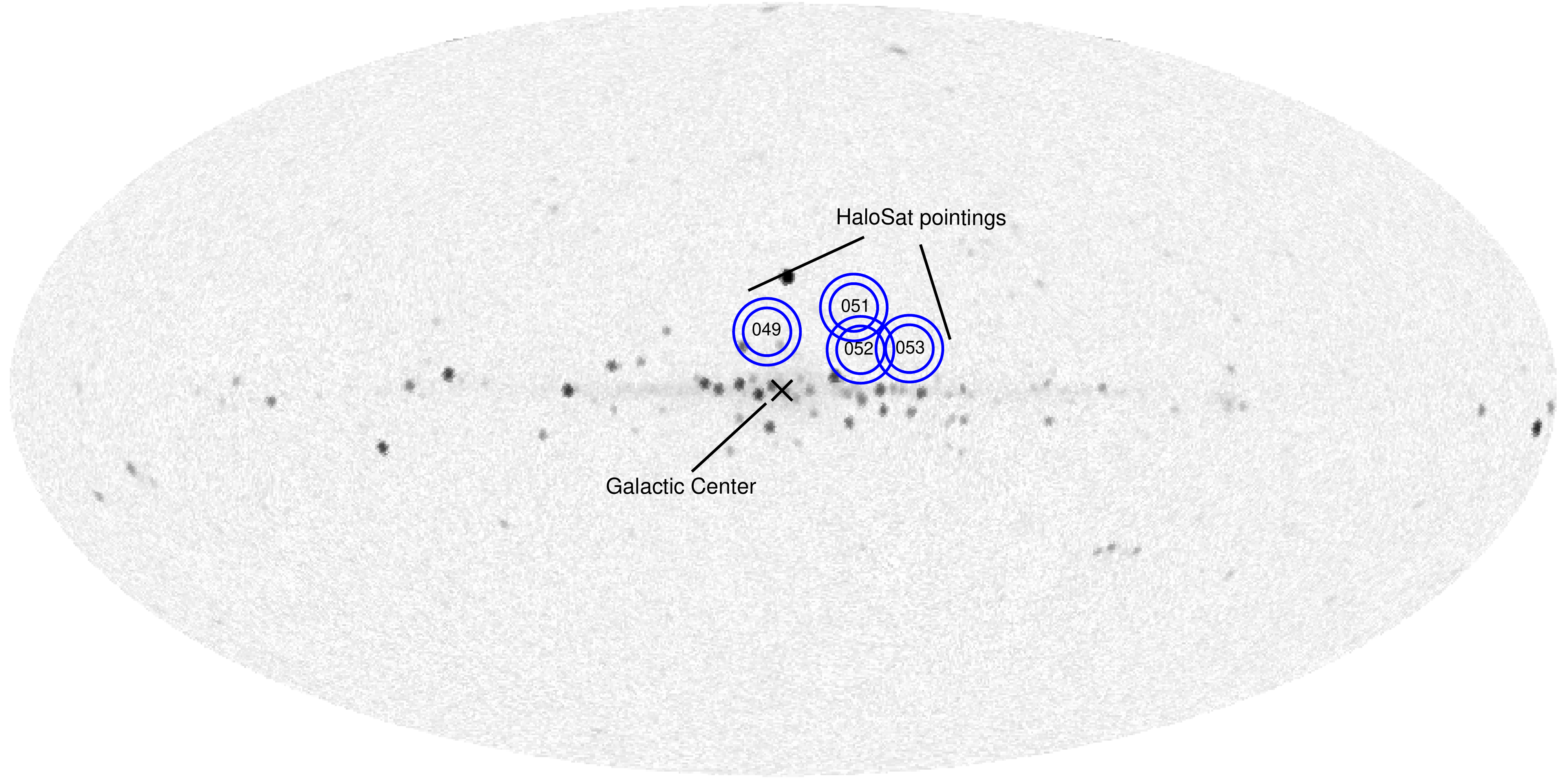}
\caption{HaloSat observations used for analysis (blue) indicated on a MAXI SSC $2-4$ keV all-sky map \citep{2020PASJ...72...17N} centered on the Galactic Center (black). The inner edges of the annuli represent HaloSat's $5^{\circ}$ (radius) full-response field of view, and outer edges represent the $7^{\circ}$ zero-response FoV. \vspace{1em} }
\end{figure*} 

We report on a search for an emission line near 3.5 keV in the Milky Way's dark matter halo using HaloSat. The observation strategy and data reduction procedures are outlined in section 2. A description of the background analysis is in section 3, and the targeted $3-4$ keV analysis is in section 4. Results are presented in section 5, and an interpretation in the context of sterile neutrino decay and comparison to previous investigations are given in section 6. Conclusions are outlined in section 7.

\section{Observations} 
\subsection{Observation strategy}

The expected 3.5 keV signal from dark matter is strongest in observations directed towards the Galactic Center, where the Milky Way's dark matter is most concentrated. However, the Galactic Center region is contaminated by many bright X-ray sources like X-ray binary systems. For analysis, we selected four HaloSat observations spanning a range of $10^{\circ} - 30^{\circ}$ from the Galactic Center with $l < 20^{\circ}$ that were relatively free from bright X-ray sources that would contaminate the diffuse emission. A summary of the coordinates, distance from the Galactic Center, and exposure time after data processing for each of the observations is given in Table 1, and the observations are indicated on a MAXI $2-4$ keV all-sky map in Figure 1. Observations were collected from 2019 May to 2020 July. 

The closest HaloSat observations to the Galactic Center are HS049 and HS052, which have the highest expected flux from a 3.5 keV line. However, HS049 is contaminated by two X-ray sources within the FoV that elevate the background contributions (see section 3.2.1). We identified HS052 as the most sensitive HaloSat observation to a 3.5 keV line, due to its proximity to the Galactic Center and absence of bright X-ray sources. Additional HaloSat observation time was allotted for HS052, and the total cleaned exposure time of $\sim$236 ks makes HS052 one of the deepest HaloSat science observations. Therefore, we expect the best constraints on a 3.5 keV line feature to be from HS052. 

\begin{deluxetable}{cccccc}[h]
\caption{Summary of HaloSat observations near the Galactic Center used for analysis}
\tablehead{\colhead{Target ID} & \colhead{$l$} & \colhead{} & \colhead{$b$} & \colhead{GC distance} & \colhead{Exposure time} \\ {} & {[deg]} & {} & {[deg]} & {[deg]} & {[ks]}}
\startdata
HS049 & 3.2 & & 12.3 & 12.7 & 183 \\
HS052 & 343.4 & & 8.5 & 18.6 & 236 \\
HS051 & 344.4 & & 17.4 & 23.2 & 96 \\
HS053 & 333.0 & & 8.7 & 28.3 & 128 
\enddata
\end{deluxetable} 

\subsection{Data reduction}
Each HaloSat observation consists of data from 3 nominally identical detectors. The detectors are specified by their corresponding data-processing unit (DPU) identification number (named DPU 14, DPU 54, and DPU 38). Data for each DPU were processed for each observation in order to remove intervals with elevated background contributions. Cuts were applied to the `hard' band ($3-7$ keV) rate and the `VLE' ($E > 7$ keV) rate in $64$ s time bins to reduce the time-variable background contributions. The hard band cut restricted count rates to be $\leq 0.16$ counts sec$^{-1}$ in that band, and the VLE band cut restricted count rates to be $\leq 0.75$ counts sec$^{-1}$ in that band. The observations have a combined exposure time of $\sim 642$ ks after all cuts were applied.

The HaloSat on-ground and on-orbit calibrations are described in \citet{2020JATIS...6d4005Z} and \citet{2019ApJ...884..162K}, respectively. 
\section{Background Analysis} 
Due to the small expected flux of the 3.5 keV line, it was necessary to obtain a physically-motivated background description in the vicinity of the line. HaloSat spectra are relatively simple above 2 keV, so we fit background models over the $2-7$ keV energy range in order to obtain a comprehensive description of the continuum. All HaloSat spectra are fit using Xspec 12.10.1f \citep{1996ASPC..101...17A} within PyXspec 2.0.2 \citep{2016HEAD...1511502A} using the C-statistic \citep{1979ApJ...228..939C} as the fit statistic.

Background models for our observations consist of contributions from two main components: the HaloSat instrumental background and an astrophysical background. We performed a simultaneous fit of the instrumental and astrophysical background models to the data from all DPUs for each observation.  

\subsection{Instrumental background}
The instrumental background for each DPU was modelled by a power law normalized by the $3-7$ keV flux (pegpwrlw) folded through a diagonal response matrix. The pegpwrlw photon index was modelled separately for each DPU as a linear function of hard band rate determined from high latitude data (Methods section of \cite{2020NatAs...4.1072K}), and the normalization was left as a free parameter. The instrumental background parameters for each observation are given in Table 2, and the process for modelling the HaloSat instrumental background is documented at the HEASARC\footnote{https://heasarc.gsfc.nasa.gov/docs/halosat/analysis/}. 

\begin{deluxetable}{ccc}[b]
\vspace{2mm}
\caption{Instrumental background pegpwrlw photon indices  }
\tablehead{ \colhead{Observation} & \colhead{DPU ID} &  \colhead{Photon index}}
\startdata
\underline{HS049} & 14 & $0.72 \pm 0.07$  \\
& 38 & $0.65 \pm 0.08$ \\
& 54 & $0.65 \pm 0.07$  \\
\underline{HS052} & 14 & $0.86 \pm 0.06$  \\
& 38 & $0.82 \pm 0.07$ \\
& 54 & $0.83 \pm 0.06$ \\
\underline{HS051} & 14 & $0.86 \pm 0.06$ \\
& 38 & $0.83 \pm 0.07$ \\
& 54 & $0.83 \pm 0.07$  \\
\underline{HS053} & 14 & $0.90 \pm 0.06$  \\
& 38 & $0.87 \pm 0.07$  \\
& 54 & $0.87 \pm 0.06$  \\
\enddata
%updated for fixed instback phind
\end{deluxetable} 

Checks were made to ensure that the spacecraft positioning (roll angle and orientation relative to the nearby bright X-ray source Sco X-1) did not affect the behaviour of the instrumental background. 

\subsection{Astrophysical background}
The astrophysical background model for each HaloSat observation above 2 keV consists of contributions from the Cosmic X-ray Background (CXB) and X-ray sources within each FoV, as required. The astrophysical background models were folded through the standard HaloSat response matrices current as of 2019 April 23. 

Contributions from unresolved extragalactic X-ray sources to the CXB were modelled by a single absorbed power law with photon index $\Gamma = 1.45$ and a 1 keV normalization of 10.91 keV cm$^{-2}$ s$^{-1}$ sr$^{-1}$ keV$^{-1}$ \citep{2017ApJ...837...19C} subject to a Galactic interstellar absorption column density calculated as the response-weighted equivalent $N_H$ across the HaloSat FoV from the SFD dust map \citep{1998ApJ...500..525S}. The calculation of the  response-weighted equivalent $N_H$ is described in \citet{2020ApJ...904...54L}. Absorptions were modelled with TBabs \citep{2000ApJ...542..914W} and fixed in spectral analysis. 

\subsubsection{HS049 source contamination}
The HS049 observation is the nearest of our observations to the Galactic Center, and thus has the highest expected flux from a 3.5 keV line. However, there are two X-ray sources within HS049 that contribute substantially to the $3-4$ keV energy range of interest: the Ophiuchus galaxy cluster and the low mass X-ray binary GX 9+9. Contributions from the Ophiuchus galaxy cluster and GX 9+9 in HS049 were evaluated using data from the Gas Slit Camera (GSC; $2-30$ keV) in the MAXI experiment 
%co-
located on the International Space Station. MAXI was operating at the time of HaloSat observations, and scans the sky daily. The MAXI data were obtained from the HEASARC archive for the period from 2019 August $3 - 2020$ July 25. Source spectra, background spectra, and response matrices were calculated using the `mxproduct' tool distributed with HEASoft 6.28 \citep{1995ASPC...77..367B}. MAXI spectra were accumulated over a period of HaloSat observations, which is justified since the spectra do not show variability.

\begin{figure}[t]
    \epsscale{1.1}
    \plotone{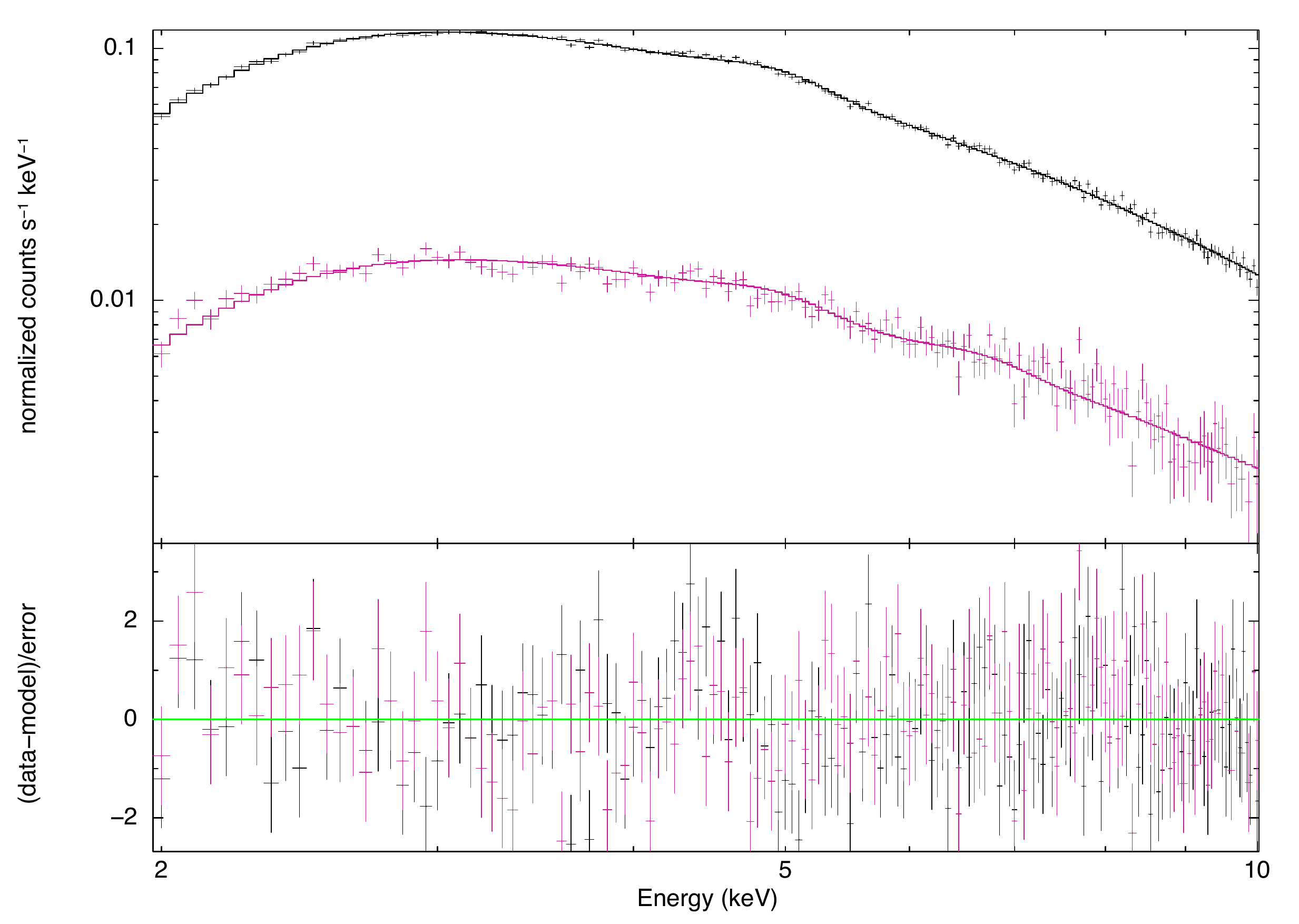}
\caption{GX 9+9 (black) and Ophiuchus galaxy cluster (magenta) spectra from MAXI, fit from 2-10 keV as HS049 background model components.}
\end{figure}

\begin{figure*}
    \vspace{-1em}
    \epsscale{1.1}
    \plotone{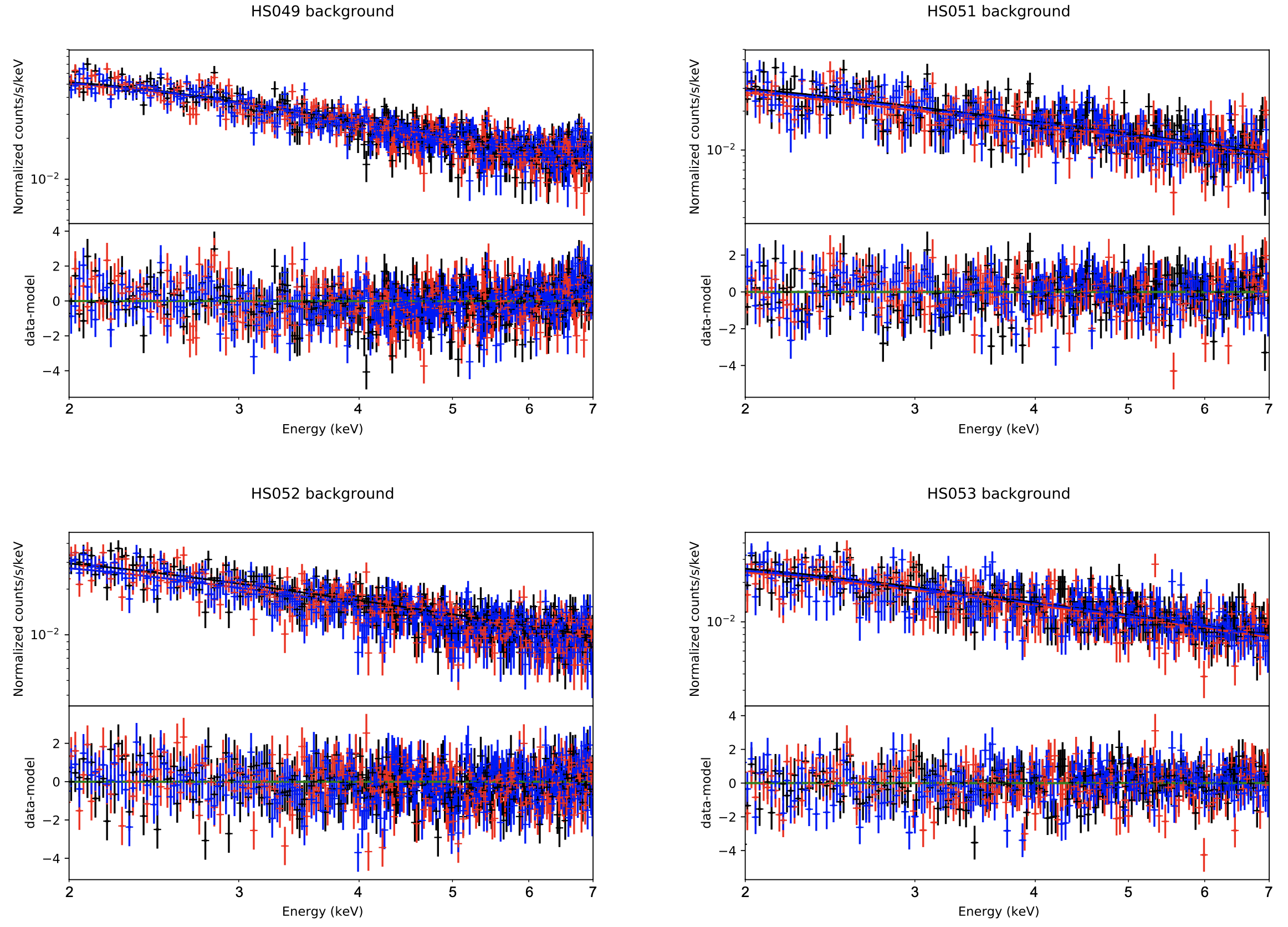}
\caption{Total $2-7$ keV background models for selected observations, with data from DPU 14 (black), DPU 38 (red), DPU 54 (blue). Each model contains contributions from the CXB and HaloSat instrumental background, while HS049 (top left) contains additional components for X-ray source contamination. \vspace{1em}} 
\end{figure*}

We performed a simultaneous fit of the MAXI spectra for the Ophiuchus galaxy cluster and GX 9+9 over the range of $2-10$ keV. Contributions from the Ophiuchus galaxy cluster were modelled with an absorbed collisionally-ionized plasma model (tbabs $\times$ apec, \citet{2008PASJ...60.1133F}). We modelled contributions from GX 9+9 with an absorbed disk-blackbody component plus a Comptonized component (tbabs $\times$ (nthcomp + diskbb), \citet{2020A&A...635A.209I}). The simultaneous fit of the models to the spectra fit the data with $\chi^2 / \text{dof} = 1.11$ (see Fig. 2).

The resulting best-fit models were used in the standard background model (CXB + instrumental background) for HS049 as fixed components to comprise the total HS049 $2-7$ keV background model. While the Ophiuchus galaxy cluster is within the $5^{\circ}$ full-response FoV, GX 9+9 is within HaloSat's partial-response FoV. So, GX 9+9 contributions were weighted by a factor which was determined by the $2-7$ keV fit in order to properly model the observed flux of GX 9+9. The total $2-7$ keV background models for each observation are shown in Figure 3. 

\section{3-4 keV Spectral Analysis} 

After obtaining background descriptions from the broad $2-7$ keV fits for each observation, we narrowed the energy range of interest to $3-4$ keV in order to investigate the presence of the 3.5 keV line while retaining an understanding of the continuum near 3.5 keV. In Xspec, the astrophysical background model components were fixed to the values determined in the $2-7$ keV fit, and the instrumental background normalization was initially set to the value from the $2-7$ keV fit and was allowed to vary.

\subsection{3-4 keV emission lines}
\begin{figure*}[t]
    \vspace{-1em}
    \epsscale{1.1}
    \plotone{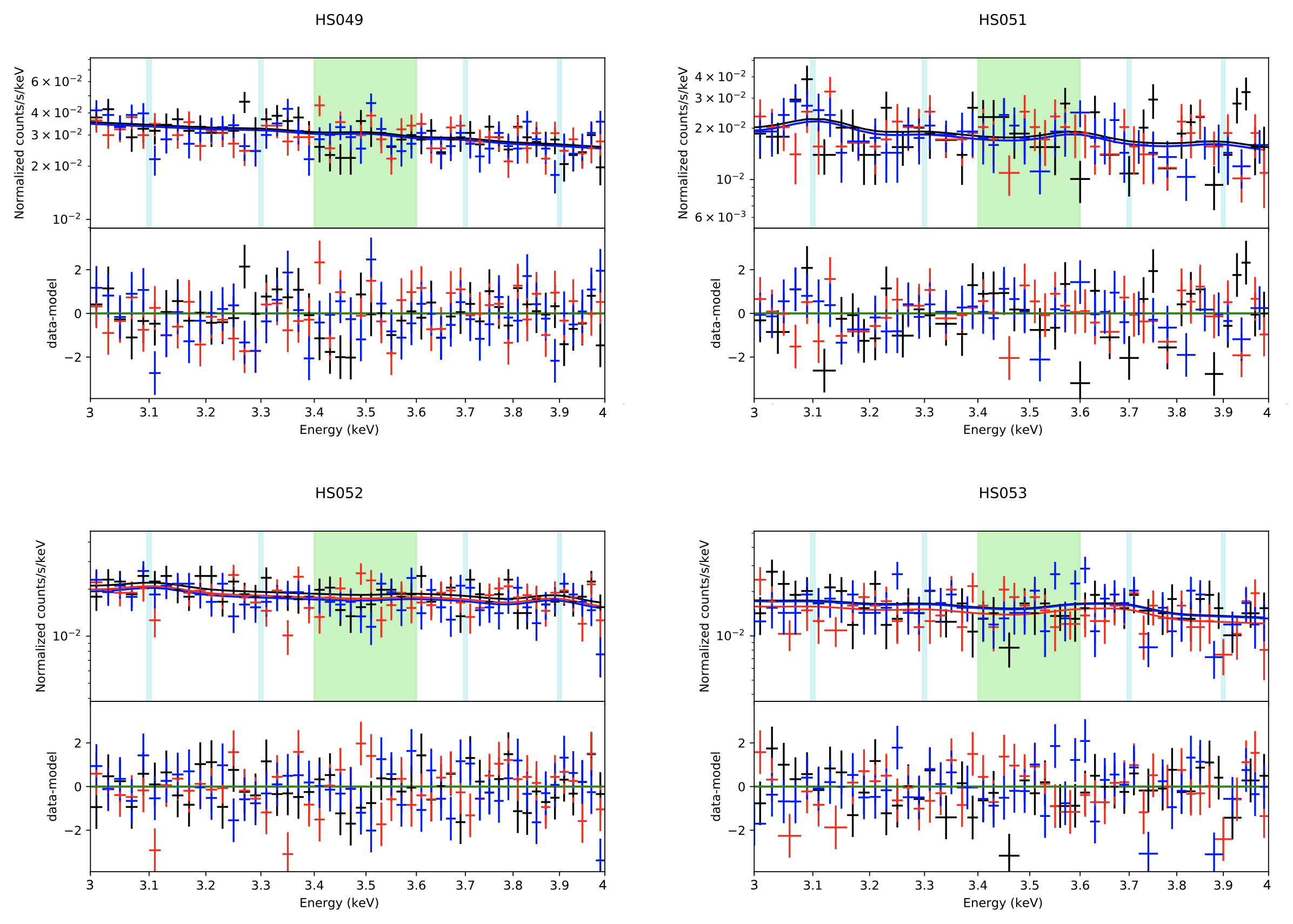}
\caption{$3-4$ keV models for selected observations, with data from DPU 14 (black), DPU 38 (red), DPU 54 (blue). Each model contains contributions from the total background in each observation, the 3.5 keV line, and additional astrophysical lines in the $3-4$ keV energy range. Blue vertical lines represent the energies at which astrophysical emission line components were added, and the green shaded region represents the $\sim 3.5 \pm 0.1$ keV range of previously reported sterile neutrino line features. \vspace{1em}} 
\end{figure*}

Previous searches for the 3.5 keV line in the Milky Way's dark matter halo using \textit{Chandra} and \textit{XMM-Newton} have included additional faint, highly ionized emission lines that could be produced in hot plasmas within galaxy clusters and the Milky Way in the $3-4$ keV energy range in their models (e.g. \citet{2015PhRvL.115p1301B}; \citet{2018arXiv181210488B}; \citet{2020ApJ...905..146S}). Recent analyses have shown that properly accounting for these faint emission lines is necessary to place physically-motivated constraints on a 3.5 keV line (see \citet{2020arXiv200406170A} and \citet{2020arXiv200406601B}). 

Emission lines from $3-4$ keV include an Ar XVII complex at $\sim$3.1 keV (\citet{2018arXiv181210488B}; \citet{2015PhRvL.115p1301B}), a Ca XIX complex at  $\sim$3.9 keV (\citet{2018arXiv181210488B}; \citet{2015PhRvL.115p1301B}), a line at 3.68 keV due either to an Ar XVII complex and/or a Ca K$\alpha$ instrumental line in \textit{Chandra} and \textit{XMM-Newton} (\citet{2018arXiv181210488B}; \citet{2020ApJ...905..146S}; \citet{2015PhRvL.115p1301B}), and a blend of Ar XVIII and S XVI lines with a K K$\alpha$ instrumental line in \textit{Chandra} and \textit{XMM-Newton} at $\sim$3.3 keV (\citet{2018arXiv181210488B}; \citet{2020ApJ...905..146S}; \citet{2015PhRvL.115p1301B}).

We added gaussian model components at fixed energies 3.1, 3.3, 3.7, and 3.9 keV with fixed narrow widths and free normalizations. Previous detections of a line near 3.5 keV have reported best-fit energies ranging from $\sim3.48 - 3.57$ keV, so we added a gaussian component with fixed narrow width, free normalization, and an energy allowed to vary from $3.4-3.6$ keV. A preliminary fit was performed for each observation in Xspec. The corresponding correlation matrix from the fit was used as the source for the covariance information for a Markov Chain Monte Carlo (MCMC) run in Xspec, from which the best-fit parameters were determined. The MCMC procedure used the Goodman-Weare (G-W; \citet{2010CAMCS...5...65G}) algorithm with 50 walkers and a chain length of $5 \times 10^6$ to ensure convergence. The best-fit instrumental background $3-4$ keV fluxes from the MCMC run were converted to fluxes under a 3.5 keV line by scaling the flux for each DPU by the FWHM of the HaloSat response at 3.5 keV. The best-fit parameters for the MCMC runs in each observation are given in Table 4, spectral fits are shown in Figure 4, and contour plots are given in the Appendix.

The fitting process was repeated for each observation after removing the gaussian component at 3.5 keV so the models could be compared and the significance associated with a 3.5 keV line feature determined. 

\subsection{Statistical methods}
A standard method of comparing the goodness-of-fit between two models is to calculate an F-statistic from the $\chi^2$ values and number of degrees of freedom corresponding to each fit. However, \citet{2002ApJ...571..545P} points out that the F-statistic is not appropriate for testing for the presence of an emission line. This is primarily because the F-test requires that the null values of the parameters being compared not be on the boundary of the allowed parameter values, and since an emission line flux can only be non-negative, the null values for the flux parameter of an emission line (i.e. zero flux) are on the boundary of possible values. Instead, \citet{2002ApJ...571..545P} recommends using a Bayesian approach to testing for the presence of an emission line. 

We use the Bayesian Information Criterion (BIC; \citet{1978AnSta...6..461S}) to evaluate the goodness-of-fit of our models. The BIC is advantageous in that it offers a way to evaluate evidence either against or in favor of the simpler of two models.

The BIC is defined as 
\begin{equation}
    \text{BIC} = -2\ln(L) + k\ln(n)
\end{equation}
for $k$ the number of parameters, $n$ the number of data points, and $L$ the maximized value of the likelihood function. The BIC tends to favor simpler models, implementing a penalty term for models with a greater number of parameters. 

The $\Delta$BIC test provides an estimate of evidence against or in favor of the simpler of two models (see Table 3, \citet{kass}). The test favors the model with the lowest BIC. The $\Delta$BIC test statistic can be calculated from Xspec's MCMC chains, when the chain is run after a preliminary C-statistic fit. We used the $\Delta$BIC test to determine the evidence for or against the presence of a 3.5 keV line in our models (see the Appendix for more details). \vspace{-1em}

\begin{deluxetable}{cc}[h]
\caption{Qualitative interpretation of the $\Delta$BIC test.  }
\tablehead{ \colhead{$\Delta$BIC} &  \colhead{Evidence for model} \\ \colhead{ } & \colhead{ with lower BIC}}
\startdata
$0 < \Delta\text{BIC} \leq 2$ & weak \\ 
$2 < \Delta\text{BIC} \leq 6$ & moderate \\ 
$6 < \Delta\text{BIC} \leq 10$ & strong \\ 
$\Delta\text{BIC} > 10$ & very strong 
\enddata
\end{deluxetable}  \pagebreak

\begin{deluxetable*}{lccccccccc}[h]
\caption{Best-fit parameters from the $3-4$ keV MCMC runs$^{**}$}
\tablehead{ \colhead{Observation} & \colhead{3.5 keV line} & \colhead{3.5 keV} & \colhead{3.3 keV} & \colhead{3.7 keV} & \colhead{3.1 keV} & \colhead{3.9 keV} & \colhead{DPU 14} & \colhead{DPU 38} & \colhead{DPU 54} \\ \colhead{} & \colhead{energy [keV]} & \colhead{line flux$^\dagger$} & \colhead{line flux$^\dagger$} & \colhead{line flux$^\dagger$} & \colhead{line flux$^\dagger$} & \colhead{line flux$^\dagger$} & \colhead{BKG flux$^{*}$} & \colhead{BKG flux$^{*}$} & \colhead{BKG flux$^{*}$}}
\startdata
HS049 & $3.52_{-0.09}^{+0.06}$ & $3.7_{-2.9}^{+9.6}$ & $2.0_{-1.1}^{+12}$ & $2.0_{-1.2}^{+11}$ & $0.1_{-0.0}^{+8.3}$ & $0.6_{-0.1}^{+10}$ & $1.8_{-0.7}^{+0.2}$ & $1.6_{-0.7}^{+0.2}$ & $1.6_{-0.7}^{+0.2}$ \\ 
HS052 & $3.59_{-0.16}^{+0.01}$ & $2.3_{-2.1}^{+5.4}$ & $0.1_{-0.0}^{+5.6}$ & $2.1_{-1.4}^{+7.4}$ & $5.2_{-3.7}^{+7.2}$ & $5.0_{-3.2}^{+6.3}$ & $3.5_{-0.4}^{+0.2}$ & $3.2_{-0.4}^{+0.2}$ & $3.0_{-0.4}^{+0.2}$ \\ 
HS051 & $3.59_{-0.18}^{+0.01}$ & $6.3_{-4.9}^{+10}$ & $2.5_{-1.7}^{+12}$ & $0.2_{-0.0}^{+11}$ & $10.3_{-5.5}^{+13}$ & $2.8_{-2.1}^{+11}$ & $3.2_{-0.7}^{+0.3}$ & $3.0_{-0.7}^{+0.3}$ & $3.0_{-0.7}^{+0.3}$ \\ 
HS053 & $3.58_{-0.15}^{+0.01}$ & $4.3_{-3.6}^{+7.6}$ & $1.9_{-1.5}^{+8.1}$ & $5.5_{-3.2}^{+9.2}$ & $1.5_{-1.1}^{+8.6}$ & $0.1_{-0.0}^{+6.3}$ & $2.5_{-0.5}^{+0.2}$ & $2.0_{-0.6}^{+0.2}$ & $2.4_{-0.6}^{+0.2}$
\enddata
\tablenotetext{}{$^\dagger$line fluxes are given in units of $10^{-2}$ ph cm$^{-2}$ s$^{-1}$ sr$^{-1}$ } \vspace{-1mm}
\tablenotetext{}{$^*$instrumental background (BKG) pegpwrlw fluxes are given in units of $10^{-2}$ ph cm$^{-2}$ s$^{-1}$ sr$^{-1}$ under the 3.5 keV line} \vspace{-1mm}
\tablenotetext{}{$^{**}$the full $3-4$ keV models include fixed components for the astrophysical background (including HS049 source contamination) } \vspace{-3em}
\end{deluxetable*} 

\section{Results} 
\begin{deluxetable*}{lccccc}[t]
\caption{$\Delta$BIC test results for selected observations.  }
\tablehead{ \colhead{Observation} &  \colhead{$\Delta$BIC} & \colhead{Evidence for model} & \colhead{Bayes} & \colhead{Probability of} & \colhead{Line} \\ \colhead{ } & \colhead{ } & \colhead{without 3.5 keV line} & \colhead{factor} & \colhead{line detection} & \colhead{significance}}
\startdata
HS049 & 9.36 & strong & 107.96 & 0.009 & $< 1\sigma$ \\ 
HS052 & 9.42 & strong & 111.29 & 0.009 & $< 1\sigma$ \\ 
HS051 & 8.58 & strong & 73.08 & 0.014 & $< 1\sigma$ \\ 
HS053 & 8.81 & strong & 81.67 & 0.012 & $< 1\sigma$ 
\enddata
\end{deluxetable*}

The $\Delta$BIC test for each observation gives strong evidence in favor of models without a line near 3.5 keV, with line significances in all observations less than $1\sigma$ (see Table 5). We report a non-detection of a line near 3.5 keV in each observation, and provide 90\% confidence level (CL) upper limits on the flux of a 3.5 keV line in Table 6.

A standard parameter of sterile neutrino models is the mixing angle, which describes the strength of the coupling between sterile and active neutrinos. The mixing angle is independent of the source observed, and provides a way to directly compare previous detections of the line in galaxy clusters and higher-latitude observations of the Milky Way's dark matter halo with the HaloSat values. We translated 90\% CL upper limits on the flux of the 3.5 keV line to corresponding 90\% CL upper limits on the mixing angle for 7.1 keV sterile neutrino dark matter. 

Many previous studies of the 3.5 keV line have assumed that the Milky Way's dark matter distribution is well-represented by a Navarro–Frenk–White (NFW) profile \citep{1996ApJ...462..563N, 1997ApJ...490..493N}. The NFW profile is given by \vspace{-1em}

\begin{eqnarray}
\begin{aligned}
    \rho_{\text{NFW}}(r) &= \frac{\rho_0 r_s^3}{r(r+r_s)^2} \\
    &= \frac{M_{200}}{4\pi R^3_{200}}\; \frac{c^3}{\ln(1+c)-\frac{c}{1+c}}\; \frac{r_s^3}{r(r+\frac{R_{200}}{c})^2}
\end{aligned}
\end{eqnarray} \textcolor{black}{as a function of the distance  $r$ from the Galactic Center, where $\rho_0$ is the characteristic density, $r_s = R_{200}/c$ is the scale radius for $R_{200}$ characterizing a sphere with mean enclosed density equal to $200 \rho_{\text{crit}}$ (for $\rho_{\text{crit}} = $ the critical density of the Universe), $M_{200}$ is the mass contained within $R_{200}$, and $c$ is the halo concentration.} 

\textcolor{black}{However, \citet{2020MNRAS.494.4291C} recently showed that the Milky Way's Galactic rotation curve is better described by an NFW profile that is modified by the Galactic baryonic distribution. While many previous descriptions of the Milky Way's dark matter density profile have neglected the contraction of the dark matter halo density caused by the accretion and settling of baryons in the Milky Way, the contracted NFW profile is motivated by predictions of hydrodynamical simulations (e.g. \citet{2015MNRAS.451.1247S, 2016MNRAS.461.2658D}) and is preferred to the uncontracted NFW profile (3) by a variety of independent observations (see \citet{2020MNRAS.494.4291C} for a full description). } 

\textcolor{black}{Therefore, in order to convert the HaloSat 3.5 keV line flux upper limits to upper limits on the 7.1 keV sterile neutrino dark matter mixing angle, we used the publicly-available \citet{2020zndo...3740067C} Python code to calculate the contracted NFW profile. The Milky Way's dark matter distribution was evaluated with the \citet{2020MNRAS.494.4291C} best-fit contracted NFW halo, which was fit from the \textit{Gaia} DR2 rotation curve as measured by \citet{2019ApJ...871..120E}. For this model, the relevant parameters are the dark matter halo mass $M_{200} = 9.7 \times 10^{11} \text{ M}_\odot$, halo concentration $c = 9.4$, and $R_{200} = 218$ kpc \citep{2020MNRAS.494.4291C}. The distance from the Sun to the Galactic Center used in the calculation is $r_\odot = 8.122$ kpc \citep{2018A&A...615L..15G}. }

The flux of a decaying dark matter line from the Milky Way is proportional to the mass of dark matter within the FoV \citep{2001ApJ...562..593A}. The flux is calculated as in \citet{2016PhRvD..93f3518N} and \citet{2020ApJ...905..146S}:
\begin{equation}
    F_{DM} = \frac{\Gamma_\gamma}{4\pi m_s} \left(\frac{M_{DM}^{FOV}}{D^2}\right)
\end{equation}
with $m_s = 7.1$ keV equal to twice the mean best-fit energy from the $3-4$ keV fits, $D$ the distance from Earth to the dark matter, and $M_{DM}^{FOV}$ the total dark matter mass within the FoV found by integrating the \textcolor{black}{\citet{2020MNRAS.494.4291C} contracted} NFW profile along a given line of sight and over the HaloSat response-weighted effective FoV.

Combining (1) and (4), the mixing angle term is solved for as 
\begin{equation}
    \sin^2(2\theta) = C  \cdot \frac{F_{DM}}{m_s^4}   \left(\frac{M_{DM}^{FOV}}{D^2}\right)^{-1}
\end{equation}
for $F_{DM}$ the flux of the 3.5 keV line and the conversion constant given by
\begin{equation}
   C = 4\pi \cdot (7.25 \times 10^{21})\; \rm{s}\; \rm{keV}^5
\end{equation}

Upper limits on the 3.5 keV line flux and the corresponding mixing angle for each observation are given in Table 6. The upper limit constraints on the flux and mixing angle range from $F \leq (0.08-0.16)$ ph cm$^{-2}$ s$^{-1}$ sr$^{-1}$ and \textcolor{black}{$\sin^2(2\theta) \leq (4.25-11) \times 10^{-11}$} across the four HaloSat observations. The upper limit derived in each observation is an independent measurement, and as described in section 2.1, HS052 is the most sensitive observation to a 3.5 keV emission line, from which we expect the best constraints. Therefore, our best constraints are from HS052, and we take our best upper limits as $F \leq 0.077$ ph cm$^{-2}$ s$^{-1}$ sr$^{-1}$ and \textcolor{black}{$\sin^2(2\theta) \leq 4.25 \times 10^{-11}$}.

\begin{deluxetable}{ccc}[h]
\caption{Upper limits derived from fits.  }
\tablehead{ \colhead{Observation} &  \colhead{3.5 keV line flux} & \colhead{$\sin^2(2\theta)$} \\ \colhead{ } &  \colhead{[ph cm$^{-2}$ s$^{-1}$ sr$^{-1}$]} & \colhead{[$10^{-11}$]}}
\startdata
HS049 & 0.133 & \textcolor{black}{5.33} \\ 
HS052 & 0.077 & \textcolor{black}{4.25} \\ 
HS051 & 0.163 & \textcolor{black}{11.0} \\ 
HS053 & 0.119 & \textcolor{black}{9.76}
\enddata 
\end{deluxetable} \vspace{-2em}

The Milky Way's dark matter distribution has been modeled in previous analyses (e.g. \citet{2018ApJ...854..179C, 2020Sci...367.1465D, 2020ApJ...905..146S}) with a simple, uncontracted NFW profile with the best-fit parameters from \citet{2013JCAP...07..016N}: the scale radius is $r_s = 16.1$ kpc and the dark matter density scale $\rho_0$ is fixed so the local dark matter density $\rho_{\text{local}} = 0.47$ GeV cm$^{-3}$ at a distance from the Sun to the Galactic Center $r_\odot = 8.08$ kpc. In order to provide a direct comparison to these analyses, we also report the best HaloSat 7.1 keV sterile neutrino dark matter mixing angle upper limit when adopting the simpler \citet{2013JCAP...07..016N} uncontracted NFW profile description of the Milky Way's dark matter distribution as $\sin^2(2\theta) \leq 3.72 \times 10^{-11}$. \textcolor{black}{However, for this model, the halo mass is $M_{200} = 1.5 \times 10^{12} \text{ M}_\odot$ \citep{2013JCAP...07..016N}, which is larger than the halo mass $M_{200} = 9.7 \times 10^{11} \text{ M}_\odot$ in the \citet{2020MNRAS.494.4291C} contracted NFW profile that we used to calculate the best HaloSat 7.1 keV sterile neutrino mixing angle upper limit. So, the the mixing angle upper limit derived with the \citet{2013JCAP...07..016N} uncontracted NFW halo model, which has a larger Milky Way dark matter halo mass, is stronger than the best HaloSat limit derived with the \citet{2020MNRAS.494.4291C} contracted NFW halo model.}

\section{Discussion} 
\subsection{Mixing angle comparisons}

The best HaloSat upper limit on the 7.1 keV sterile neutrino mixing angle is \textcolor{black}{$\sin^2(2\theta) \leq 4.25 \times 10^{-11}$}. This constraint is consistent with two recent mixing angle estimates from 3.5 keV line detections and upper limits in the Milky Way's dark matter halo \citep{2018arXiv181210488B, 2020ApJ...905..146S}. The \citet{2018arXiv181210488B} mixing angle $\sin^2(2\theta) = (1.6-2.1) \times 10^{-11}$ derived from an \textit{XMM-Newton} line detection in the Milky Way's dark matter halo is contained by our upper limit. The \citet{2020ApJ...905..146S} mixing angle upper limit derived from \textit{Chandra} observations of the Milky Way's dark matter halo is $\sin^2(2\theta) \leq  2.58 \times 10^{-11}$, which is potentially increased by a factor of $\sim$2 by systematic uncertainties related to the choice of parameters \textcolor{black}{for the uncontracted NFW profile model of the Milky Way's dark matter distribution}. The increased upper limit $\sin^2(2\theta) \leq  4.96 \times 10^{-11}$ is higher than the mixing angle upper limit found in this work, while the stricter upper limit $\sin^2(2\theta) \leq  2.58 \times 10^{-11}$ is contained by our upper limit. Thus, the HaloSat upper limit is consistent with the \citet{2018arXiv181210488B} detection and \citet{2020ApJ...905..146S} upper limit. Figure 5 compares the HaloSat upper limits with the \citet{2018arXiv181210488B} detection and \citet{2020ApJ...905..146S} upper limit in terms of the 3.5 keV line flux corresponding to the reported mixing angles. The HaloSat analysis has systematics, observations, and analysis methods that are different from the \textit{Chandra} and \textit{XMM-Newton} analyses, and still provides independently-derived upper limits that are consistent with these previous analyses, which do not exclude the sterile neutrino decay interpretation of the 3.5 keV line feature.

\begin{figure}[t]
    \epsscale{1.15}
    \plotone{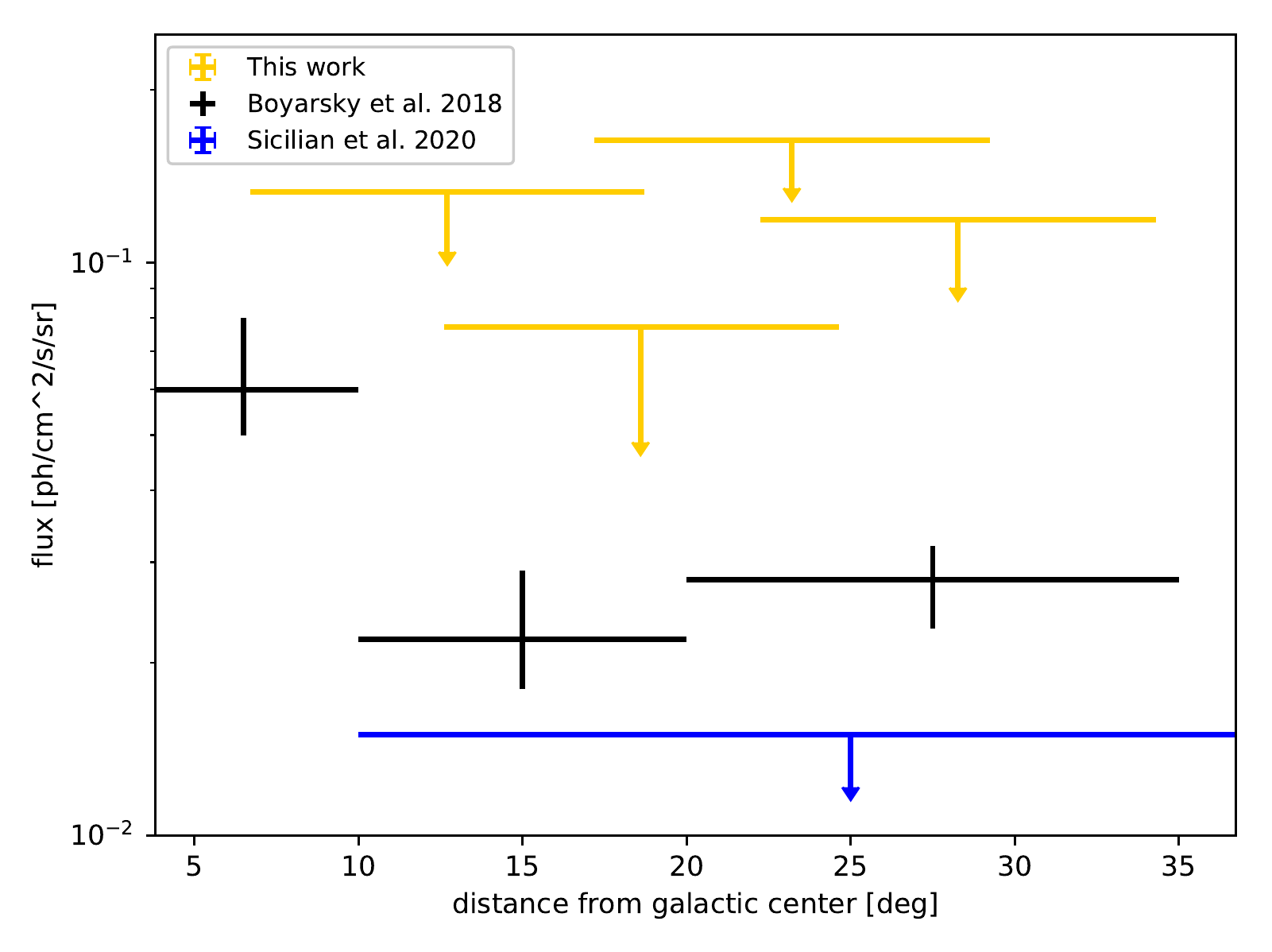}
\caption{3.5 keV line flux upper limits from this work compared with \citet{2018arXiv181210488B} line detections and \citet{2020ApJ...905..146S} upper limits in the Milky Way Galactic Center region, as a function of angular distance from the Galactic Center.} 
\end{figure}

The mixing angle quoted by \citet{2018ApJ...854..179C} for the detection of a line near 3.5 keV in \textit{Chandra} observations of the Milky Way's dark matter halo of $\sin^2(2\theta) = (0.83-2.75) \times 10^{-10}$ is strongly excluded by our upper limit. However, \citet{2018ApJ...854..179C} were unable to rule out a statistical fluctuation as the cause of the 3.5 keV line. Therefore, when we treat the \citet{2018ApJ...854..179C} mixing angle as a $3\sigma$ upper limit, the HaloSat upper limit places a strong constraint on the \citet{2018ApJ...854..179C} limit by approximately a factor of 5. Therefore it is unlikely that the 3.5 keV line feature reported in \citet{2018ApJ...854..179C} is associated with the decay of $\sim$7 keV sterile neutrino dark matter.

A $\sim$30 Ms \textit{XMM-Newton} blank-sky analysis reported a non-detection of a line near 3.5 keV and estimated a mixing angle upper limit of $\sin^2(2\theta) \sim 10^{-12}$ that was inconsistent with the sterile neutrino decay interpretation \citep{2020Sci...367.1465D}. The \citet{2020Sci...367.1465D} result was recently questioned primarily due to the treatment of the astrophysical background, the energy range over which spectra were fit ($3.3-3.8$ keV), and the Milky Way dark matter density profile parameters assumed. \citet{2020arXiv200406601B} shows that by modelling additional astrophysical emission lines at 3.1, 3.3, 3.7, and 3.9 keV and fitting over a broader energy range of $3-4$ keV to constrain the continuum, the \citet{2020Sci...367.1465D} limit is weakened by more than an order of magnitude. \citet{2020arXiv200406170A} showed that accounting for the model dependencies, including the dark matter density profile parameters assumed and the exclusion of astrophysical emission lines at 3.3 and 3.7 keV, relax the \citet{2020Sci...367.1465D} upper limit by at least a factor of $\sim$20. Our HaloSat mixing angle upper limit is in better agreement with the \citet{2020arXiv200406601B} and \citet{2020arXiv200406170A} analyses, with a similar analysis method in that we allow for the presence of the four known emission lines from $3-4$ keV. The HaloSat analysis does not prefer the original \citet{2020Sci...367.1465D} limit over the subsequent revised limits in \citet{2020arXiv200406601B} and \citet{2020arXiv200406170A} since the HaloSat upper limit is over an order of magnitude higher than the \citet{2020Sci...367.1465D} limit.

The HaloSat mixing angle upper limit places constraints on a number of detections and upper limits derived from galaxy cluster spectra. In particular, the original \citet{2014ApJ...789...13B} mixing angle derived from an \textit{XMM-Newton} line detection in 73 galaxy clusters at $\sin^2(2\theta) \sim 7 \times 10^{-11}$ is excluded by the HaloSat upper limit. The lower bound of the \citet{2014PhRvL.113y1301B} mixing angle range $\sin^2(2\theta) = (2-20) \times 10^{-11}$ derived from \textit{XMM-Newton} observations of the M31 galaxy and the Perseus galaxy cluster is consistent with the HaloSat upper limit, while the higher bound is strongly constrained. A \textit{Suzaku} analysis of 47 galaxy clusters reported an upper limit at $\sin^2(2\theta) \leq 6.1 \times 10^{-11}$ \citep{2016ApJ...831...55B}, which is further constrained by our limit. A recent \textit{XMM-Newton} galaxy cluster analysis reported the mixing angle upper limit $\sin^2(2\theta) \leq 4.4 \times 10^{-11}$ \citep{2020MNRAS.497..656B}, which is only marginally higher than the HaloSat upper limit. 

The mixing angle upper limit derived in this work is inconsistent with the \citet{2018ApJ...854..179C} and \citet{2014ApJ...789...13B} line detections made with \textit{Chandra} and \textit{XMM-Newton}, but is consistent with the \citet{2014PhRvL.113y1301B} and \citet{2018arXiv181210488B} detections with \textit{XMM-Newton}. The mixing angle upper limit further constrains a number of previous estimates derived from observations of the Milky Way's dark matter halo and galaxy clusters. \citet{2017ApJ...837L..15A} points out the inconsistency between the \textit{Hitomi} non-detection in the Perseus cluster and the \citet{2014ApJ...789...13B} detection with the \textit{XMM-Newton} MOS in the Perseus cluster is attributable to a systematic error in the \textit{XMM-Newton} result. The HaloSat limit cannot rule out the sterile neutrino decay interpretation of the line, but also cannot rule out any potential instrumental effects in previous detections of the line with CCDs.

\subsection{Strong evidence criteria}
Given that we did not significantly detect a line at $\sim$3.5 keV in any observation, we investigated how strong the 3.5 keV line would have to be in order for the fitting procedure to show strong evidence in favor of models with a line near 3.5 keV. In Xspec, we simulated spectra (using the fakeit command) based on the best-fit models from the $3-4$ keV fits for each observation, modifying the strength of the 3.5 keV gaussian line component for each simulation. We stepped over a grid from $0.001 - 0.01$ ph cm$^{-2}$ s$^{-1}$ with 15 linearly-spaced values, which were used as the Xspec normalization of a 3.5 keV line in the simulated spectra. The simulated spectra were fit with the same fitting procedure as the $3-4$ keV fits. We found that the 3.5 keV line would need to have a strength of $\sim$0.16 ph cm$^{-2}$ s$^{-1}$ sr$^{-1}$ to provide strong evidence in favor of models with a line near 3.5 keV (see Fig. 6). This estimate is consistent with the high end of our flux upper limits for each HaloSat observation (see Table 6). 

\begin{figure}[b]
    \epsscale{1.1}
    %\vspace{-1em}
    \plotone{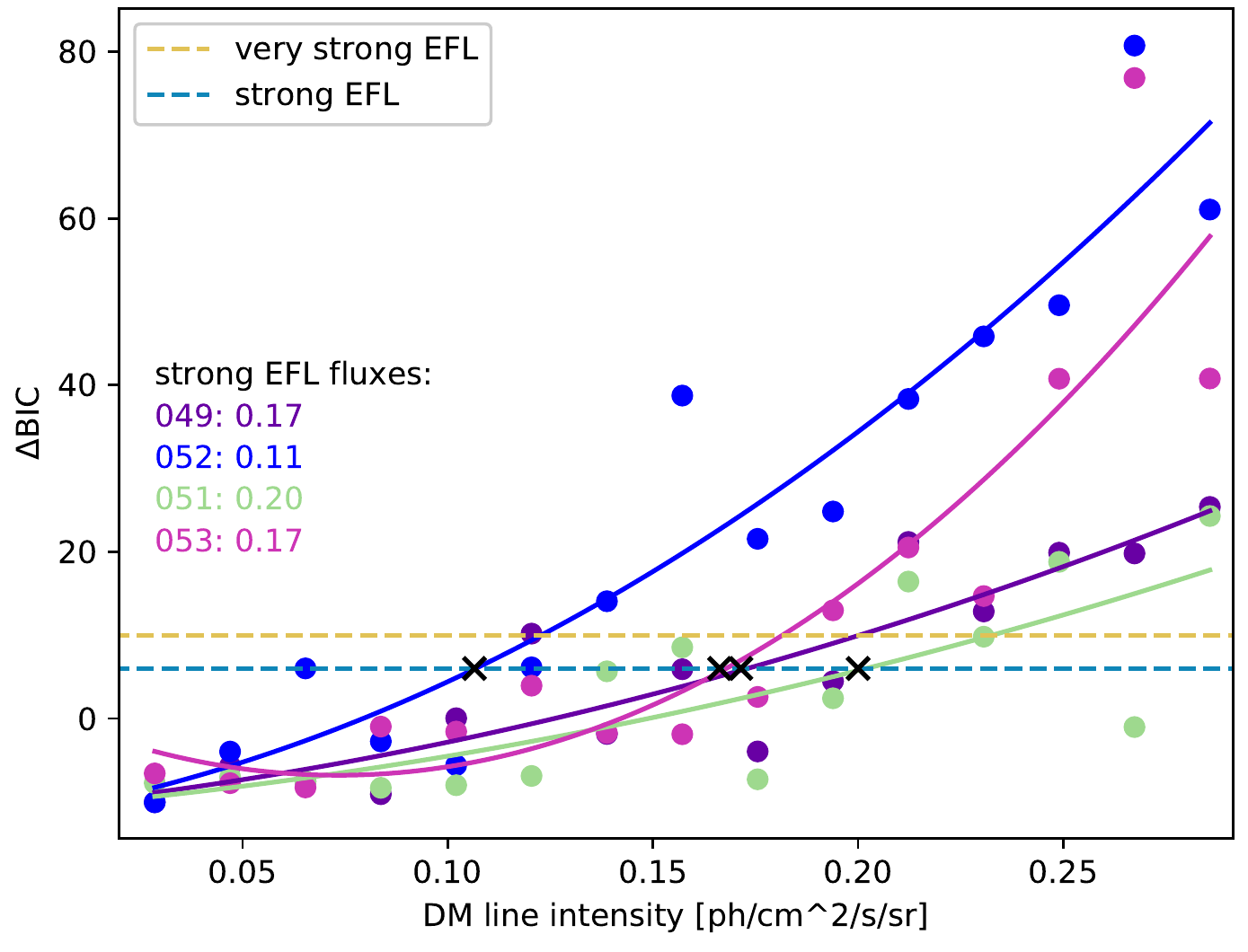}
\caption{$\Delta$BIC evidence for simulated spectra, indicating at which 3.5 keV line strength our fitting procedure shows strong evidence (EFL) for models with a line near 3.5 keV.}
\end{figure}

We also point out that one should be cautious when comparing reported values of the 3.5 keV line significance, which can depend strongly on the choice of statistics. Previous analyses have utilized a $\chi^2$ approach to test for the presence of a line near 3.5 keV. In order to understand the consequences of this statistical approach, we repeated the statistical analysis using the $\Delta \chi^2$ test to compare models. In this case, the $\Delta$BIC test tended to prefer simpler models with fewer parameters more strongly than a $\Delta \chi^2$ method, but the $\Delta \chi^2$ test also yields non-detections of a 3.5 keV line, with significances \textcolor{black}{of detections} consistently $<1\sigma$.

\section{Conclusions} 
We report a non-detection of an emission line feature near 3.5 keV in four HaloSat observations of the Milky Way's dark matter halo from $10^{\circ} - 30^{\circ}$ of the Galactic Center using the BIC to determine the significance of the line feature, \textcolor{black}{and assuming the Milky Way's dark matter distribution is well-described by the \citet{2020MNRAS.494.4291C} contracted NFW profile}. The HaloSat 90\% CL upper limit on the 3.5 keV line flux is $0.077$ ph cm$^{-2}$ s$^{-1}$ sr$^{-1}$, which corresponds to a 7.1 keV sterile neutrino mixing angle upper limit of \textcolor{black}{$\sin^2(2\theta) \leq 4.25 \times 10^{-11}$}. The HaloSat mixing angle upper limit is consistent with the \citet{2014PhRvL.113y1301B} and \citet{2018arXiv181210488B} 3.5 keV line detection with \textit{XMM-Newton} but excludes the \citet{2018ApJ...854..179C} and \citet{2014ApJ...789...13B} line detections made with \textit{Chandra} and \textit{XMM-Newton}, while placing constraints on a number of previous mixing angle estimates derived from observations of the Milky Way's dark matter halo and galaxy clusters. Based on mixing angle upper limits, the HaloSat analysis cannot rule out possible effects inherent to the CCD instruments contributing at least in part to the previous 3.5 keV line detections. Despite placing a constraint on the 7.1 keV sterile neutrino parameter space, the sterile neutrino decay interpretation of the 3.5 keV line feature cannot be excluded by this analysis.

\acknowledgements{}
The HaloSat mission is supported by NASA grant NNX15AU57G. This project/material is based upon work supported by the Iowa Space Grant Consortium under NASA Award No. NNX16AL88H, the Goldwater Scholarship Foundation, the Iowa Center for Research by Undergraduates, and a NASA Goddard Space Flight Center summer internship. \textcolor{black}{We thank the anonymous reviewer for comments which helped improve this manuscript.} %change to thanking the referee 

\facility{HaloSat}
\software{Xspec 12.10.1f \citep{1996ASPC..101...17A}, PyXspec 2.0.2 \citep{2016HEAD...1511502A}, corner.py \citep{2016JOSS....1...24F}, HEASoft 6.28 and the MAXI FTOOLS package 27Mar2020\_V02.1 \citep{1995ASPC...77..367B}}

\appendix 
\restartappendixnumbering
\section{BIC calculation \& MCMC contour plots}

The BIC statistic for each model was calculated using the output file from an MCMC run in Xspec. We used the G-W algorithm with a given number of walkers and chain length, and covariance information taken from a preliminary fit to the data using the C-statistic as fit statistic. \\

The MCMC chain output by Xspec is a FITS file containing a set of parameters and corresponding fit statistic for the length of the chain, so for example in this analysis, the output files contained $5 \times 10^6$ entries of parameters and associated C-statistic. \\

For Poisson-distributed data, the likelihood is given as 
\begin{equation}
    L = \prod_{i=1}^{N} (tm_i)^{S_i} e^{-tm_i} / S_i!
\end{equation} for $S_i$ the observed counts, $t$ the exposure time, and $m_i$ the expected counts based on the model and instrument response. \\

The C-statistic in each entry of the output FITS file is the maximum likelihood-based statistic defined as 
\begin{equation}
\begin{split}
 C =& \;2 \sum_{i=1}^{N} (tm_i) - S_i + S_i (\ln(S_i) - \ln(tm_i)) \\
 =& -2 \; \ln(L) \\
\end{split}
\end{equation}

The best-fit parameter set is that which maximizes the likelihood (A1), which translates to minimizing the C-statistic (A2). We read the FITS file for each MCMC run into Python and searched for the parameter set corresponding to the minimum C-statistic. This is the set of best-fit parameters for the chain. \\ 

For the BIC given in (2), the minimum C-statistic corresponding to the best-fit set of parameters can by definition be input into the BIC calculation for a model: 
\begin{equation}
    \text{BIC} = \text{minimum(C-statistic)} + k\ln(n)
\end{equation}

The test statistic $\Delta$BIC is then simply the difference between BIC for two models (in this analysis, the models with and without an emission line component near 3.5 keV). The $\Delta$BIC calculation can be easily run with an addition to plot the MCMC contour plots from the MCMC output FITS file entries in order to output the statistical information and data visualizations with one procedure. Figure A1 shows the joint confidence regions for pairs of parameters for the HS052 MCMC run. 

%From the $\Delta$BIC, we estimated the corresponding Bayes factor:
%\begin{equation}
%    \text{BF} \simeq e^{(0.5 \cdot \Delta\text{BIC})}
%\end{equation}
 
\begin{figure*}[h!]
    \epsscale{1.2}
    \plotone{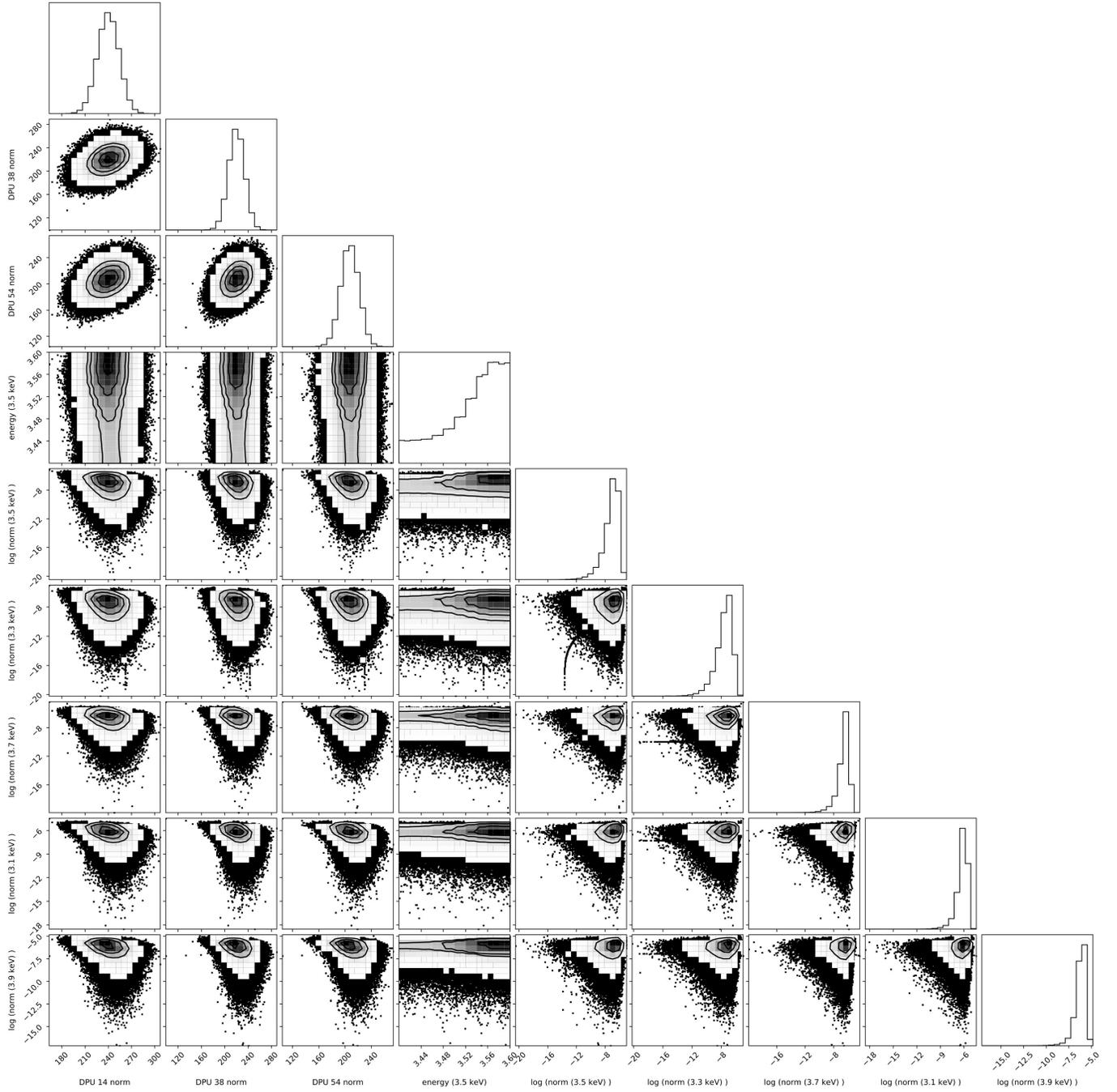}
\caption{Contour plots from the HS052 $3-4$ keV fit using the MCMC run that provided the best upper limit}
\end{figure*}

\end{document}